\documentclass[review]{elsarticle}

\usepackage{lineno,hyperref}
\modulolinenumbers[5]
\usepackage{graphicx,caption}
\usepackage{float}

\usepackage{subfigure}
\usepackage{fullpage}
\usepackage{wrapfig}

\journal{Journal of Nucelar Instrumentation and Methods}

\begin{document}

\begin{frontmatter}


\title{Operation and Performance of Microhexcavity Pixel Detector in Gas Discharge and Avalanche Mode}



\author[address1]{A. Mulski}

\author[address3]{Y. Benhammou}
\author[address1]{J.W. Chapman}
\author[address3]{A. Das}
\author[address3]{E. Etzion}
\author[address1]{C. Ferretti}
\author[address2]{P.S. Friedman}
\author[address4]{R.P. Johnson}
\author[address1]{D.S. Levin*}
\author[address1]{N. Kamp}
\author[address1]{H. Ochoa}
\author[address3]{M. Raviv-Moshe}
\author[address1]{N. Ristow}

\address[address1]{Department of Physics, University of Michigan, Ann Arbor, MI, 48109}
\address[address2]{Integrated  Sensors, LLC, 2403 Evergreen Rd., Ottawa Hills, OH, 43606}
\address[address3]{School of Physics and Astronomy, Tel Aviv University, Tel Aviv, 69978, Israel}
\address[address4]{Department of Physics, University of California Santa Cruz, Santa Cruz, CA, 95064}

\cortext[ca]{Corresponding author}


\begin{abstract}
The Microhexcavity Panel ($\mu$Hex) is a novel gaseous micropattern particle detector comprised of
a dense array of close-packed hexagonal pixels, each operating as an independent detection unit for ionizing radiation. 
It is a second generation detector derived from plasma panel detectors  and  microcavity detectors. The $\mu$Hex  is under development to be deployed  as a scalable, fast timing (ns)  and  hermetically sealed gaseous tracking detector with high rate ($\rm > 100 KHz/cm^2$) capability. The devices reported here were fabricated as $16 \times 16$ pixel arrays of  $\rm 2 \; mm$ edge-to-edge, 1 mm deep hexagonal cells embedded in a thin, 1.4 mm glass-ceramic wafer. Cell walls are metalized cathodes, connected to high voltage bus lines through conductive vias.  Anodes are small, 457 $\mu$m diameter metal discs screen printed on the upper substrate.  The detectors are filled with an operating gas to near 1 atm and then closed with a shut-off valve. They have been operated in both avalanche mode  and gas discharge devices,  producing mV to volt level signals with about 1 to 3  ns rise times.  Operation in discharge mode 
is enabled by high impedance quench resistors on the high voltage bus at each pixel site.  Results indicate that each individual pixel behaves as an isolated detection unit with  
high single pixel intrinsic efficiency to both $\beta$s from radioactive sources and to cosmic ray muons. Continuous avalanche mode operation over several days  at hit rates over 300 $\rm KHz/cm^2$ with no 
gas flow have been observed. Measurements of pixel isolation, timing response, efficiency,  hit rate and rate stability are reported. 
\end{abstract}

\end{frontmatter}

\newpage
\section{Introduction}
The Microhexcavity Panel ($\mu$Hex) is a novel, gaseous micropattern particle detector comprised of a dense array of close-packed hexagonal pixels, each operating as an independent detection unit for ionizing radiation. The $\mu$Hex  is under development as a scalable, fast timing (ns) and ultimately, sealed gas-hermetic  tracking detector with high rate ($\rm > 100 KHz/cm^2$) capability.  Derived from the earlier plasma panel detector~\cite{PPS} and  microcavity detectors~\cite{MicroCav}, they  are fabricated as $16 \times 16$ pixel arrays of  $\rm 2 \; mm$ edge-to-edge, 1 mm deep hexagonal cells embedded  in a thin 1.4 mm glass-ceramic  substrate. Cover plates can be glass or glass-ceramic,  300 $\mu$m thick. Pixel walls are  metalized cathodes, connected to high voltage bus lines through conductive vias.  Anodes are small, thick-film discs (457 $ \mu$m diameter)  screen printed on the internal surface of the top substrate, and positioned on the cavity central axis.  The detectors are filled with an operating gas to 740 Torr  and then closed with a shut-off valve. They have been operated in both a gas discharge  mode (GD)  and in a Townsend avalanche (TA)  mode.  The resulting signals  for avalanche and discharge modes have mV to volt amplitudes with  approximately 1 to 3 ns rise times.  In both operational modes  signals are initiated by ionizing radiation traversing the pixel gas volume and producing primary and secondary ion-pairs that lead to avalanche multiplication.  The GD  mode uses  a Ne + Ar Penning mixture ~\cite{Penning} with a fraction of $\rm CF_4 $ and a gas fill pressure of 740 Torr.  This mode  can be considered Geiger-like in that it  yields very large gains of order $10^9$, large amplitude volt-level signals which can be easily read out without amplification~\cite{SAULI}. The avalanche leads to streamer formation and gas breakdown that collapses the internal electrical field and terminates the discharge. An external quench resistor is used to connect each pixel to its high voltage bus line. This resistance, combined with the pixel capacitance,  sets a recovery time of the pixel. 
Resistances of 500 M$\Omega$ and 1 G$\Omega$  were used, with results from the latter included here.  Pixel capacitance is of order 0.5-1 pF, measured with  an LCR meter. The recovery time allows full neutralization of ions and prevents signal regeneration. While the single pixel recovery time is many hundreds of $\mu$sec, limiting the maximum single pixel rate,  the array of independent pixels potentially allows a higher rate capability. Operation in  TA mode is done with the GD mode quench resistors left in place. TA mode  produces small,  mV level signals that require amplification before they can be discriminated. Unlike in GD mode,  many fewer ions and electrons are formed in the avalanche process,  the electric field remains established and the pixel can respond to higher rates. The rate capability  is  ultimately  limited by space charge effects.  In TA mode the operating gas was  pure $\rm CF_4$ or  $\rm CF_4$ combined with  $\rm C_2F_6$ at 740 Torr.  Results from both GD and TA modes are reported in the following.
\par
\begin{figure}[H]\centering
  \includegraphics[width=0.75\textwidth]{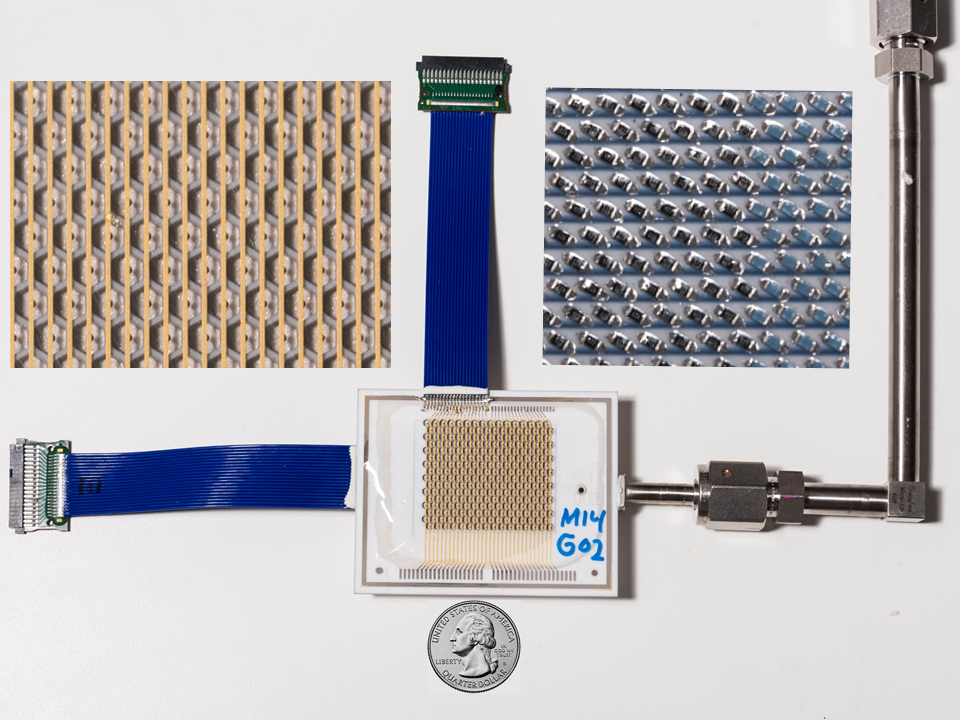}
  \vspace{-2mm}
  \caption{Assembled  detector with gas fill port, readout and high voltage feed cables. The gas feed tube  connects to a shut-off valve, not shown. Left inset: cavities with readout bus lines. Right inset: array of quench resistors. Color online.}
  \label{fig:uhex}
\end{figure}

A $\mu$Hex detector with its gas port and fill tube is shown in Figure~\ref{fig:uhex}.  Insets show the pixels and quench resistors. The pixel inset also shows the readout lines. These lines connect to the anodes through the conductive vias which are masked off in the image.  The ribbon cables are used both to convey the high voltage to each row of pixels and for read out of each column of pixels.  Preparation of the detector for operation includes baking at moderate temperatures to accelerate outgassing of impurities, extensive pump down and evacuation of the fill tube and cavity volumes, and backfilling with the particular operating gas. The gas is prepared {\it in situ} by a gas mixing station using mass flow controllers. The gas mixing precision is estimated to be $\pm5\%$ in the  fraction of any gas component. The large fill tube provides a gas reservoir of many tens of cc's in addition to the small gas volume. After backfilling of the desired gas mixture  a permanent valve (not shown in Figure~\ref{fig:uhex}) on the fill tube is closed. In this condition, detectors have been operated in both modes months after a fill has occurred.  In GD mode readout is done by extracting the signal from  termination resistors on each readout line and routing them to discriminators and  to a multi-channel scaler or  attenuated and sent to a front-end TDC/ADC system~\cite{MDTelx}  that has been repurposed from another application.  In TA mode  signals are either delivered to standalone amplifiers and discriminated or read out directly (unattenuated) by the same TDC/ADC system.

\section{ Experimental results}

The  performance of the $\mu$Hex was evaluated initially in GD mode.  Experiments reported here addressed pixel efficiency and pixel isolation. Subsequently  it was determined that TA mode offered higher rate capability, and further testing was done in this mode. For both modes of operation a field of 128 pixels (of which 125 were known to have functioning high voltage connections) was instrumented for readout using 1G$\Omega$ quench resistors on each pixel. In GD mode, the detector was filled with the Ne-based gas and operated over a range from -900 to -1100 V.  Experiments and measurements  were conducted using a $\beta$ radiation source and cosmic ray muons. Reported below are results for pixel isolation and the efficiency to detect near-minimum ionizing  cosmic ray muons.  In TA mode measurements of rates, rate stability and time resolution were conducted and herein reported. 

\subsection{ Pixel isolation}
\begin{figure}[htpb]
 \begin{minipage}{0.48\linewidth}
    \centering
  \includegraphics[width=1.2\textwidth]{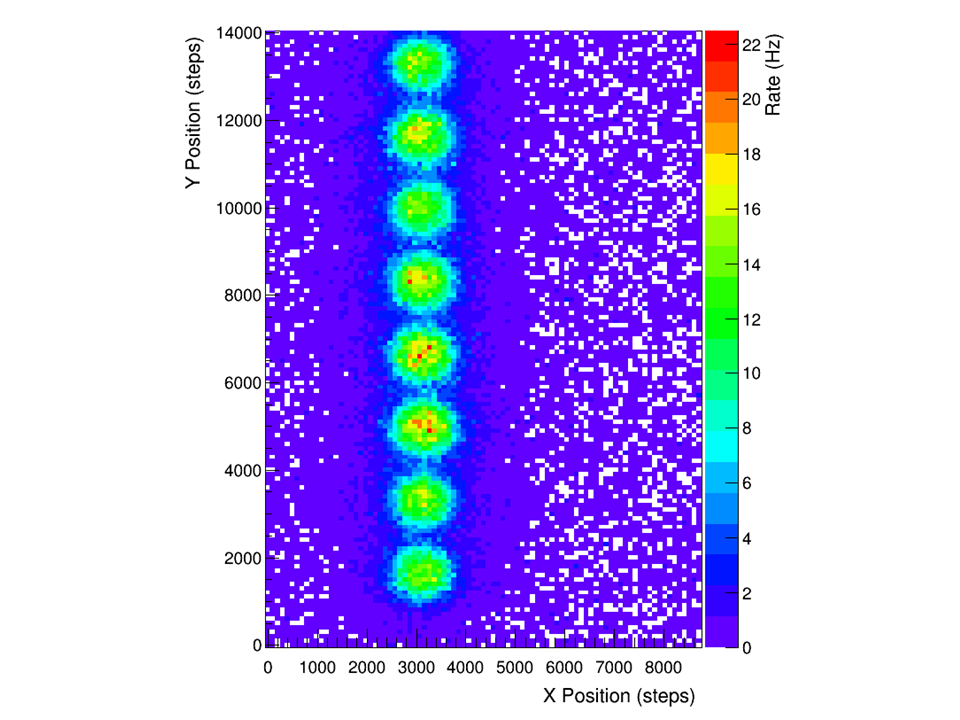}
     \caption{ Hit rate as a function of the XY position of a 1 mm collimated $\beta$ flux scanning over the  detector surface. Data from one representative  readout line is shown.   One ``step" unit on the X or Y axis  corresponds to 4 $\mu$m steps of the stepper motor. The 9000 step range is 3.6 cm.  Color online. }
  \label{fig:IsolationOne}
    \end{minipage}
    \hfill 
    \begin{minipage}{0.48\linewidth}
    \centering
    \vskip -5mm
 \includegraphics[width=1.2\textwidth]{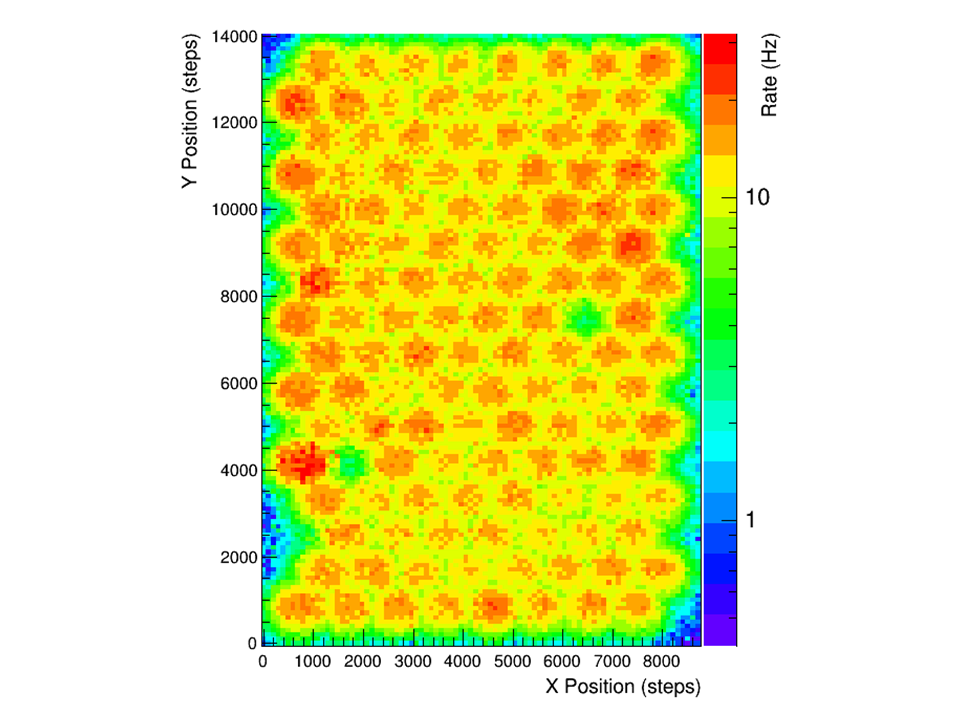}
 \vskip 0mm
  \caption{Hit rate as a function of the XY position of a 1 mm collimated beam  scanning over the detector surface. Data from all readout lines are shown.  One ``step" unit as desribed in Figure~\ref{fig:IsolationOne}. Color online.}
  \label{fig:IsolationAll}
 \end{minipage}
\end{figure}

As noted above,  $\mu$Hex pixels are designed to operate as independent, isolated detection units, an intended result of the metalized pixel cavity walls which provide both electrical and optical isolation. 
The pixel isolation was investigated in GD mode by irradiating the detector with  a $\rm ^{90}Sr$  $\beta$ source, collimated to 1 mm diameter. The source was robotically translated in  $\rm 250 \ \mu$m increments over the full  XY coordinate detector area, with the collimator aperture transverse to the plane of the detector. Sixteen readout lines of eight pixels per line were read out using a simple discriminator and multi-channel scaler. Figure~\ref{fig:IsolationOne}  
shows the hit distributions obtained for a single readout line when the whole pixel array is active. This plot demonstrates that each pixel generates an approximately equal hit rate response when exposed to the  collimated $\beta$s.  Furthermore, when the $\beta$ flux is off of the pixel associated to the readout line in Figure~\ref{fig:IsolationOne}, but positioned on any of the other pixels, this readout line registers very few hits, which were strays from the source. The background rates for off-pixel collimator  positions were  much less than 1 Hz, and the background rate when no source is near the detector was negligible.   A similar plot in Figure~\ref{fig:IsolationAll}  displays the hit rates of {\it all} readout lines combined in the scan and shows, qualitatively, the 125 active pixel boundaries. Noticeable are also three hole or zero rate regions of inactive pixels to which the high voltage bus lines were not connected.

\subsection { Cosmic ray muon sensitivity}
The response of the $\mu$Hex to cosmic ray muons  was measured in  GD mode.  Sea level cosmic ray muons have an average energy of 1-4 GeV~\cite{PDG} and are near-minimum ionizing.  The detector efficiency is likely to be lower  in GD mode, which uses a Ne-majority gas, than in TA mode which uses mostly $\rm CF_4$,  as the Ne  gas yields over four times fewer  primary ion-pairs per unit track length 
through the pixel gas volume~\cite{Sharma}. Detection of cosmic muons was done using an experimental configuration that consisted of the $\mu$Hex positioned between two paddles of a plastic scintillator hodoscope.  The pixel array comprised of 125 working pixels  was instrumented for this measurement.  An efficiency, $\epsilon$, was determined from: 
 $ \left( \frac{N_{coinc}}{N_{scint}}\right)_{exp} = \epsilon \times  \left (\frac{N_{coinc}}{N_{scint}}\right)_{MC}$
where $N_{scint}$ is the number of hodoscope hits, $N_{coinc}$ is the number of $\mu$Hex plus hodoscope coincident hits within a 1$\mu$s time window. The left side of the equation is  experimentally measured  and the right side 
is estimated using a GEANT4~\cite{G4} Monte Carlo (MC) simulation that models the hodoscope and $\mu$Hex detector geometry, the cosmic muon $cos^2(\theta)$ angular  distribution~\cite{PDG},  and the hodoscope spatial non-uniformities. The efficiency is  plotted in Figure~\ref{fig:eff} as a function of the high voltage.  The error bars are determined by the statistical errors and  the uncertainty in the geometrical configuration and scintillator efficiencies.  
A plateau from -1000 to -1060 V indicates  an operating region of maximum efficiency.  The vertical axis is the total pixel efficiency.  
A fit to the plateau gives  $\epsilon = 0.58\pm 0.01$ (fit error).  This reported efficiency is lower than unity  because the MC generates all possible track lengths through the pixel, including near zero-length tracks that clip the corners of the pixel gas volume. 
The short track lengths of near-minimum ionizing particles in the Ne-based gas, have  less than unity Poisson probability to generate a primary ion pair necessary to initiate a signal.  The measurement setup, in contrast to the MC,  did not allow to constrain the tracks to near the vertical axis or specify a minimal track length through the pixel volume.   The shaded region represents an estimated  maximum single pixel efficiency based on  the MC. It  includes  the track length distribution, the specific gas primary ion-pair production parameters, the muon sea-level energy spectrum~\cite{MUSPEC}  and the  angular distribution, and the Poisson probability that  at least  one primary  ion-pair  is produced by the track.  The  width of this region represents the combined systematic uncertainties  of these parameters.


\subsection {Hit rate and stability}
 
\begin{figure}[htpb]
 \begin{minipage}{0.48\linewidth}
    \centering
    \includegraphics[width=1.0\textwidth]{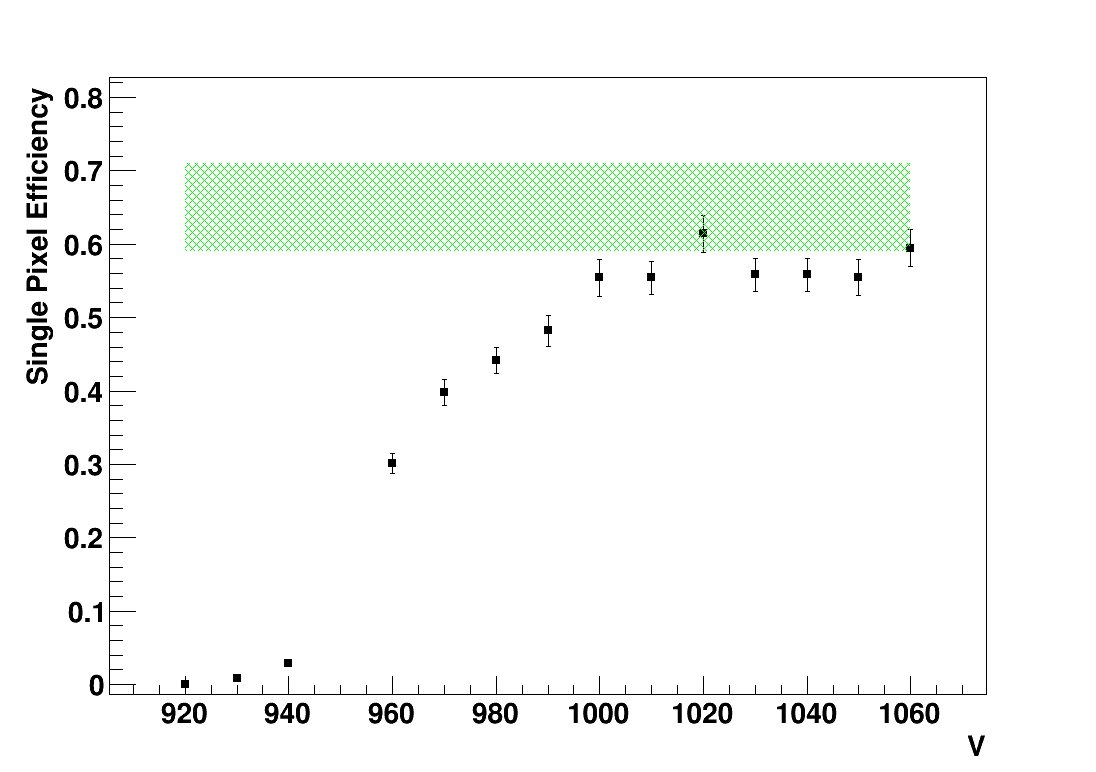}
    \vskip -3mm
  \caption{ Single pixel efficiency curve for cosmic ray muons.  Error bars are statistical. The shaded region is the estimated  maximum single pixel efficiency with systematic uncertainty determined from the MC. 
  The horizontal scale is the applied negative high voltage. Color online.}
  \label{fig:eff}
    \end{minipage}
    \hfill 
    \begin{minipage}{0.48\linewidth}
    \centering
\vskip -10mm
 \includegraphics[width=1.0\textwidth]{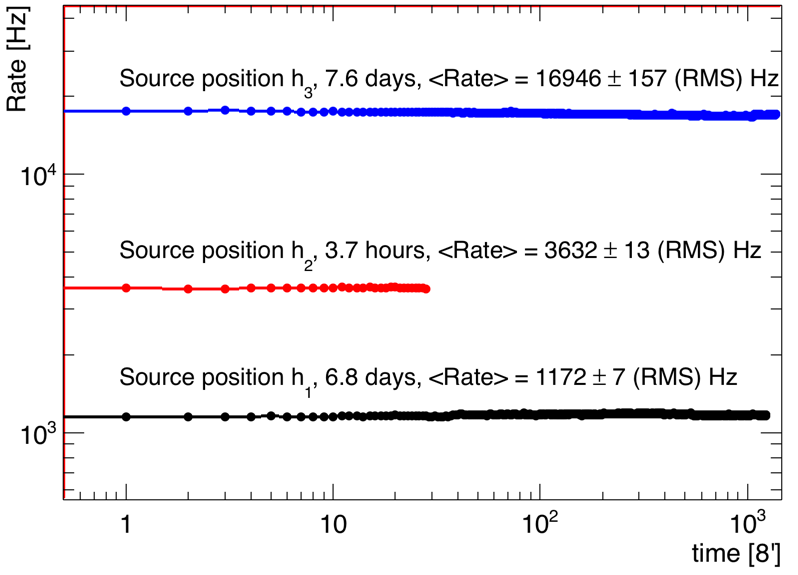}
    \vskip -5mm    
  \caption{ Single pixel signal rates  in TA mode  recorded using $^{90}$Sr $\beta$ source at three positions.  The horizontal axis has 8 minute wide bins. Color online.}
  \label{fig:hitrate3pos}
 \end{minipage}
\end{figure}

The single pixel hit rates were measured in  TA mode. The gas fill was majority  $\rm CF_4$ with  $\rm C_2F_6$ at 740 Torr. With this gas, operating voltages extended from -2800 to -3400 V. A $^{90}$Sr $\beta$ source was  positioned over the pixel.  Three source  positions were used: $\rm h_1$, $\rm h_2$ and $\rm h_3$ at approximately 3 cm, 1.5 cm  above, or directly on the pixel, respectively.  Data were acquired at position $\rm h_1$ for a  period of 6.8 days, and at positions $\rm h_2$ for 3.7  hours and $\rm h_3$ for 7.6 days. The hit rates for each  position  are reported in Figure~\ref{fig:hitrate3pos}, where each data point corresponds to the average  rate measured in an eight minute time window. As the source is moved closer to the pixel the hit rates increased  as expected.
\par
The  measured single pixel rate at position $h_1$ was  $1172 \pm 7$ (RMS) Hz and was stable to 0.6\%  over the seven day period.  Some drift of this rate of order $\pm10$ Hz over the course of the measurement  was attributed to  small variations in background hits, and to drift in the  readout electronics discrimination threshold as well. These rate fluctuations were  not associated with the detector response.  The rate at position $h_2$ was  $3632\pm 13$ (RMS)  Hz and at position $h_3$ was  $16946 \pm 157$ (RMS)  Hz.   At this last position there was an initial diminishment of the measured rate of about 3\% in the first 40 hours. For the remaining period of data taking (about 143 hours)  the rate was within  0.5\% of the average. When these  rates are normalized to the total $\rm 4.8 mm^2$ pixel area (which includes the insensitive cell walls) they  correspond to about $\rm 24 \ KHz/cm^2$, $\rm 74 \ KHz/cm^2$ and $\rm 346 \rm \ KHz/cm^2$, respectively.  The highest rate produced in this experiment was limited by the source strength and does not imply the maximum rate capability of a $\mu$Hex pixel.

\subsection {Time response}
\begin{figure}[H]\centering
  \includegraphics[width=0.6\textwidth]{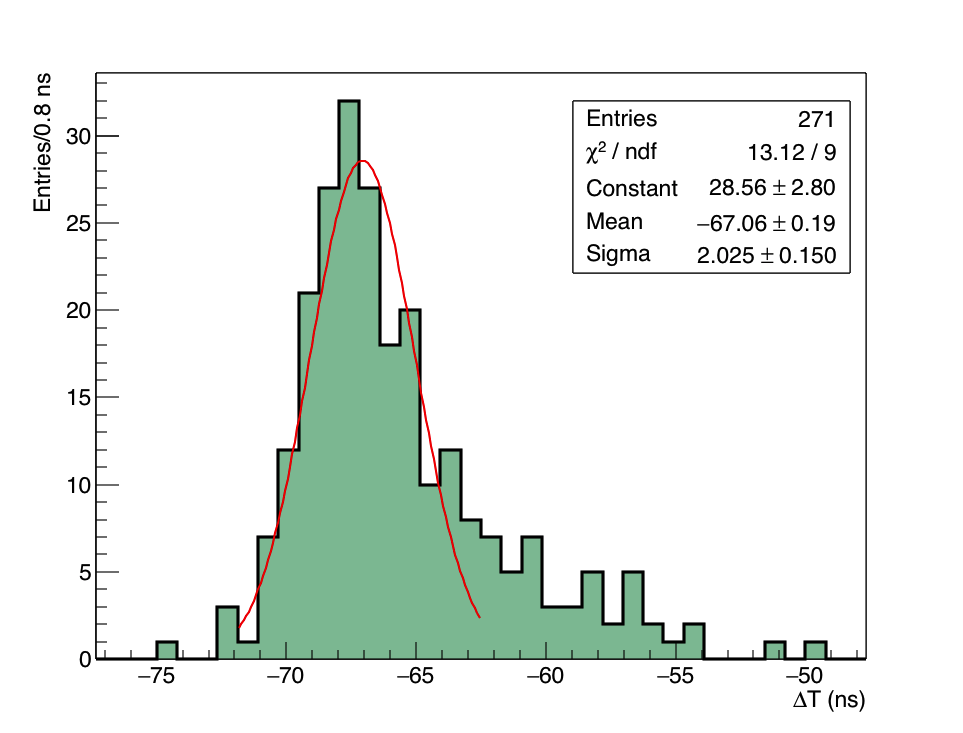}
  \vspace{-5mm}
  \caption{Arrival time distribution of cosmic ray muon hits in a $\mu$Hex with respect to the trigger time. The offset is from arbitrary cable and electronic delays. The reported Gaussian parameters are for a fit restricted to the central core of the TDC distribution. Color online.}
  \label{fig:TDCspectrum}
\end{figure}

The timing response using the $\mu$Hex in TA mode  was measured using the same scintillator hodoscope and similar geometrical configuration used to detect cosmic ray muons. An array of 128 pixels, of which 125 pixels were  operational, was used  for this measurement.   The gas fill was 100\%  $\rm CF_4$ at 740 Torr.  The high voltage was set to -3000 V.   Data were acquired  Time-to-Digital (TDC) system providing  0.78125 ns timing resolution. The  hodoscope trigger provided the reference time with an intrinsic  0.9 ns jitter.   Figure~\ref{fig:TDCspectrum}  shows the 
arrival time TDC spectrum formed from the $\mu$Hex recorded hit time subtracted from the hodoscope  trigger time. The arbitrary offset is due to various delay constants from cables and  electronics.  This TDC spectrum represents drift times from hits at any point in the pixel. The narrow width of the spectrum determined from a  range-limited Gaussian fit is $2.0\pm 0.15$ ns. This width includes also 0.9 ns of trigger jitter.  Subtraction in quadrature of this jitter reduces the timing resolution to $1.8 \pm 0.15$ ns. 

\section{Summary}
Results of the $\mu$Hex, a  pixelated, closed gas detector, have been presented. This device has been evaluated as  a high gain Geiger-like gas discharge mode detector using a Ne-based gas and quench resistors at each pixel site, and 
as a lower gain detector where signals are produced from  gas avalanche in a $\rm CF_4$-majority gas. With a Ne-based gas, the detector was operated in GD mode and  had a single pixel efficiency of 58\%$\pm$1\%, near the maximum efficiency expected for the gas and pixel geometry used. Improvements in the detector efficiency are the subject of ongoing efforts and are anticipated from larger, deeper  pixels and reduced cell wall thickness, and in TA mode 
using a $\rm CF_4$-based gas.  Overall, the $\mu$Hex has been shown to function as an array of independent detection units with fast timing. In TA mode the  time response for cosmic ray muons relative to  a scintillator trigger  time is 2 ns, determined by a Gaussian fit to the core of the arrival time  spectrum. Furthermore, TA mode hit rate measurements suggest  that the detector can operate for extended periods without degradation. A hit rate of $\rm 24 KHz/cm^2$ was measured to be stable to 0.6\% (RMS) over seven days and about  $\rm 350 KHz/cm^2$ at 0.9\% RMS variation for nearly eight days of continuous operation.

\section*{Acknowledgements}

Development of the $\mu$Hex project was funded by the U.S. National Science Foundation Grant 1506117.
The research at Tel Aviv University (TAU) was supported in part by the I-CORE Program of
the Planning and Budgeting Committee, (grant no.1937/12) and the Abramson Center. 
Research at TAU and scientific exchange and  collaboration between the University of Michigan and TAU was supported by
the Israel-United States Binational Science Foundation 2014716.  Photographs of the $\mu$Hex detector were graciously 
provided by {\it Natural Portraits and Events Photography} at https://www.naturalportraitsandevents.com.


\end{document}